\documentclass[twocolumn,showpacs,preprintnumbers,amsmath,amssymb,twocolums,mathbbm]{revtex4}

\usepackage[dvips]{graphics}
\usepackage{amsmath, amsthm, amssymb}
\usepackage{graphicx}
\usepackage{times}
\usepackage{dcolumn}
\usepackage{bm}

\newtheorem{theorem}{Theorem}

\newtheorem{lemma}{Lemma}

\newcommand{\bra}[1]{\langle #1|}
\newcommand{\ket}[1]{|#1\rangle}
\newcommand{\braket}[2]{\langle #1|#2\rangle}
\newcommand{\tr}{\text{tr}}

\newcommand{\id}{\mathbb{I}}

\begin{document}

%\preprint{APS/123-QED}

\title{Entanglement activation and the robustness of quantum correlations}

\author{Fernando G.S.L. Brand\~ao}
\email{fernando.brandao@imperial.ac.uk}
\affiliation{${}^{1}$QOLS, Blackett Laboratory, Imperial College London, London SW7 2BW, UK}
\affiliation{${}^{2}$Institute for Mathematical Sciences, Imperial College London, London SW7 2BW, UK}

\date{\today}

\begin{abstract}

We show that the usefulness of a state as an activator in teleportation protocols is equivalent to the robustness of its entanglement to noise. The robustness of entanglement of a bi-partite state $\sigma$ is linked to the maximum increase in the the fidelity of teleportation of any other state when $\sigma$ is used as an extra resource. On the one hand, this connection gives an operational meaning to the robustness of entanglement. On the other hand, it shows that the activation capability - which has a central role as an operational way of quantifying bound entangled states - can be estimated experimentally by measuring entanglement witnesses. 
\end{abstract}
%\pacs{03.67.Mn}

\maketitle

\section*{}
The use of quantum mechanical systems to send and process information in a different, and sometimes more efficient, way than possible by classical means has attracted a lot of interest recently (see e.g. \cite{Nielsen}). Among many applications - such as quantum computing, dense coding, and quantum cryptography \cite{Nielsen} - quantum teleportation \cite{Ben1} stands out as a radically new manner of transmitting quantum information: by using pre-established entanglement, Alice can send a possibly unknown quantum state to Bob by merely communicating classical bits. Fundamental in this process is the  entanglement shared by the parties. As such, considerably effort has been devoted to find the precise connection between bi-partite entanglement and quantum teleportation \cite{Ben1, Ben2, Hor1, Hor2, Hor3, Hor4, Voll1, Ver1, Mas1, fiu, ad}.

In the simplest case, Alice and Bob share a single copy of a quantum state $\rho$, which is used to teleport a state from Alice to Bob. It is clear that if $\rho$ is not entangled, the best that Bob can do is to guess what was the state of Alice's system. It turns out that this is also the case for some entangled states \cite{Hor4}. For other states, better than classical teleportation is possible, although the fidelity of the teleported state with the initial one might be smaller than unit: there is a threshold for the maximal attainable fidelity of teleportation \cite{Hor4}. An interesting phenomenon in this setting is the one in which an entangled state - which by itself cannot be used to perform teleportation better than by classical means - can be employed to increase the teleportation threshold of another state \cite{Hor3, Voll1, Mas1}. In this way, the entanglement of the former state, which is said to be activated, can be used after all, although in an undirect manner. Recently, it was proven that every bi-partite entangled state, including all bound entangled states \cite{Hor2}, can be used as an activator \cite{Mas1}. 
	
In this work we address the question of how to quantify the usefulness of a bi-partite state $\sigma$ as an activator. We introduce a figure of merit which naturally quantifies the maximum increase in the fidelity of teleportation of a bi-partite state that $\sigma$ can provide when used as an extra resource. Somehow surprisingly, we find that this figure of merit turns out to be equal to the robustness of entanglement of $\sigma$ \cite{Vid1}, a well-studied entanglement measure with a clear physical interpretation as the minimum amount of separable noise necessary to completely wash out the entanglement of $\sigma$ \cite{Vid1}. Hence, it is established a quantitative relation between entanglement activation and the endurance to noise of quantum correlations, two properties of entanglement which a priori dot not seem to be connected.  

This connection has interesting consequences both to the theory of entanglement measures and to the experimental estimation of entanglement \cite{Wit}. On one side, we find an operational meaning - in the sense of quantifying the usefulness of entanglement in a particular protocol - for the robustness of entanglement \cite{Vid1}, an entanglement measure which has find many applications in the study of entanglement \cite{Vid1, Bran1, Har1, Cal1, Sash1, Stein1, QuantWit}. In particular, the robustness becomes the only known operational entanglement measure having a non-zero value on \textit{every} entangled state. On the other side, using a recently established connection \cite{Bran1, QuantWit} between the robustness of entanglement and entanglement witnesses \cite{Wit}, the expectation value of the later is given an \textit{operational} meaning: it is a lower bound to the activation capacity of the measured state. This connection implies that it is possible to derive tight bounds to the usefulness of a bi-partite state to quantum information processing even if the state in question is completely unknown. 

Consider any teleportation protocol where a bi-partite state $\rho \in {\cal D}({\cal H})$ (the set of density matrices acting on the Hilbert space ${\cal H} := {\mathbb{C}^{m} \otimes \mathbb{C}^{m}}$) is used as a $d$-dimensional quantum channel between Alice and Bob. The most general teleportation protocol that they can implement consists of several rounds of local operations and classical communication (LOCC), and can be represented as a trace preserving completely positive map $\Omega$ mapping the state to be teleported $\ket{\psi} \in \mathbb{C}^d$ held by Alice and the entangled state used as a resource $\rho$ into the final teleported state in Bob's possession. \footnote{We adopt the convention of calling a trace preserving LOCC operation simply by a LOCC map. We reserve the term SLOCC for the non trace-preserving LOCC operations.}. One can quantify the usefulness of the state $\rho$ and protocol $\Omega$ for teleporting states $\ket{\psi} \in \mathbb{C}^{d}$ using the average fidelity of the output and input states:
\begin{equation} 
f_d(\rho, \Omega) := \int d\psi
\hspace{0.1 cm} \bra{\psi}\Omega\left(\ket{\psi}\bra{\psi} \otimes \rho \right)\ket{\psi},
\end{equation}
where the integral is performed with respect to the Haar measure $d\psi$ over all pure states in $\mathbb{C}^d$ \cite{Hor4}.

In Ref. \cite{Hor4} it was shown that without loss of generality we can restrict ourselves to the class of protocols formed by the following steps: (1) a LOCC map $\Lambda: {\cal B}(\mathbb{C}^m \otimes \mathbb{C}^m) \rightarrow {\cal B}(\mathbb{C}^d \otimes \mathbb{C}^d) $ is applied to the shared state $\rho$; (2) the output state $\Lambda(\rho)$ is twirled by LOCC into a $d \times d$ isotropic state \cite{Hor4} $\rho_I$  having the same fidelity as $\Lambda(\rho)$ with the $d \times d$ maximally entangled state; (3) the standard teleportation protocol \cite{Ben1} is performed using $\rho_I$. In this case, the fidelity of teleportation is given by \cite{Hor4}
\begin{equation}
f_d(\rho, \Lambda) = \frac{F_d(\rho, \Lambda)d + 1}{d + 1},
\end{equation}
where
\begin{equation}
F_d(\rho, \Lambda) := \tr[\Lambda(\rho)\phi_d],
\end{equation}   
and $\phi_d := \ket{\phi_d}\bra{\phi_d}$, with $\ket{\phi_d} = (1/\sqrt{d})\sum_{k=1}^d \ket{d, d}$, is the projector onto the $d \times d$ maximally entangled state. It can be shown that the best fidelity achievable by a separable state is $f_{class} = 2 / (d + 1)$ \cite{Hor4}. 

In the entanglement activation setting, we have two states, $\rho$ and $\sigma$, and a LOCC map $\Lambda$ mapping $\rho \otimes \sigma$ into a $d \times d$ state. The state $\rho$, which is assumed to be finite dimensional, is such that $f_d(\rho, \Omega) \leq f_{class}$ for every LOCC map $\Omega$: it is useless for teleportation by itself. For later use, we denote the set of all such states by ${\cal M}$. A natural figure of merit of how good $\sigma$ is as an activator is how larger than the classical threshold $f_{class}$ the quantity $f_d(\rho \otimes \sigma, \Lambda)$ is. Of course, this increase strongly depends also on the LOCC map $\Lambda$ and on the state $\rho$ chosen. Hence, we consider instead the ratio of this difference with
\begin{equation}
G_d(\rho, \Lambda) := \max_{\pi \in {\cal S}} \left( f_{class} - f_d(\rho \otimes \pi, \Lambda) \right),
\end{equation}
where ${\cal S}$ denotes the set of all finite dimensional separable states. This quantity expresses how spread the fidelity is, when an arbitrary separable state is use in the place of $\sigma$, and gives an estimate of how hard it is to increase the fidelity of teleportation of the specific state $\rho$ above the classical limit.  

The activation capacity of $\sigma$ is defined as 
\begin{equation} \label{Ed}
E_d(\sigma):= \sup_{\Lambda \in LOCC} \sup_{\rho \in {\cal M}} \frac{f_d(\rho \otimes \sigma, \Lambda) - f_{class}}{G_d(\rho, \Lambda)}.
\end{equation}
That is, we are interested in the largest increase possible over all LOCC maps $\Lambda$ and states $\rho \in {\cal M}$.

The robustness of $\sigma$ relative to $\pi$, $R(\sigma||\pi)$, is defined as the smallest non-negative number $s$ such that the state
\begin{equation}   
\frac{1}{1 + s}\sigma + \frac{s}{1 + s}\pi
\end{equation}
is separable \cite{Vid1}. The robustness of entanglement of a bi-partite state $\sigma$ is given by 
\begin{equation}
R(\sigma) = \min_{\pi \in {\cal S}(\cal H)}R(\sigma||\pi).
\end{equation}
It can be interpreted as the minimum amount of separable noise necessary to wash out all the quantum correlations originally contained in $\sigma$ \cite{Vid1}. As it is the case of all other entanglement measures with a geometric character, the robustness of entanglement lacks an operational meaning in the sense of quantifying the usefulness of $\sigma$ for some quantum information task. The main result of this Letter is the following

\begin{theorem}
For every bi-partite entangled state $\sigma$ and every natural number $d \geq 2$,
\begin{equation}
E_d(\sigma) = R(\sigma)	
\end{equation}		
\end{theorem}	

For the case of two qubits, a linear relation between the maximum fidelity of teleportation and the generalized robustness of entanglement \cite{Stein1} was derived in Ref. \cite{Ver1}. Such an expression involving directly the maximum fidelity of teleportation, albeit being remarkable, cannot be extended to higher dimensional systems, as attested by the existence of bound entangled states, which have non-zero generalized robustness \cite{Vid1}, but only achieve the classical threshold for the maximum fidelity \cite{Hor4}.

\textit{Proof of Theorem 1:}

Throughout the proof we consider that $\sigma \in {\cal B}({\cal H}_{A_1} \otimes {\cal H}_{B_1})$ and $\rho \in {\cal B}({\cal H}_{A_2} \otimes {\cal H}_{A_3} \otimes {\cal H}_{B_2} \otimes {\cal H}_{B_3})$, with ${\cal H}_{A_1} = {\cal H}_{B_1} = {\cal H}_{A_2} = {\cal H}_{B_2} = \mathbb{C}^m$ and ${\cal H}_{A_3} = {\cal H}_{B_3} = \mathbb{C}^d$. We are interested in the entanglement with respect to the partition $AB$. The reason for partitioning $\rho$ in four parties will become clear in the sequel. For the sake of clarity we will break the proof in several parts. 

The first step is to rewrite Eq. (\ref{Ed}) as
\begin{equation}
E_d(\sigma) = - \inf_{\Lambda \in LOCC} \inf_{\rho \in {\cal M}} \tr(W_{\Lambda, \rho} \sigma),
\end{equation}
with
\begin{equation} \label{WWW}
W_{\Lambda, \rho} := \frac{\id - d \tr_{A_1B_1}[\id \otimes \rho \Lambda^{\cal y}(\phi_d)]}{(d + 1)G_d(\rho, \Lambda)},
\end{equation}
where $\tr_{A_1B_1}$ is the partial trace with respect to subsystems $A_1$ and $B_1$. From the fact that $\rho \in {\cal M}$, we have that $\tr(W_{\Lambda, \rho} \pi) \geq 0 $ for every separable state $\pi$. Thus $W_{\Lambda, \rho}$ is an entanglement witness. On the other hand, from the definition of $G_d(\rho, \Lambda)$, we easily find that for every separable state $\pi$, $\tr(W_{\Lambda, \rho} \pi) \leq 1$. 

It was shown in Ref. \cite{Bran1} that the robustness of entanglement can be written as
\begin{equation}
R(\sigma) = - \min_{W \in {\cal W}} \tr(W \sigma),
\end{equation}
where ${\cal W}$ is the set of all entanglement witnesses acting on ${\cal H}_{A_1} \otimes {\cal H}_{B_1}$ such that $\tr(W \pi) \leq 1$ for every separable state $\pi$. From the previous paragraph we have that $W_{\Lambda, \rho} \in {\cal W}$ and hence
\begin{equation}
E_d(\sigma) \leq R(\sigma).
\end{equation}

To complete the proof, we must show the converse inequality. In order to do so, we need the following Lemma, which is an adaptation of the results presented in Ref. \cite{Mas1}. Let ${\cal M}({\cal H})$ be the subset of ${\cal M}$ containing every state $\rho \in {\cal M}$ which acts on the Hilbert space ${\cal H} := {\cal H}_{A_2} \otimes {\cal H}_{A_3} \otimes {\cal H}_{B_2} \otimes {\cal H}_{B_3}$.

\begin{lemma}
There is a LOCC map $\Lambda$ such that for every entangled state $\sigma$, there is a $\rho \in {\cal M}({\cal H})$ such that
\begin{equation} \label{desiq}
\tr(W_{\Lambda, \rho} \sigma) < 0.
\end{equation}
\end{lemma}
\begin{proof}
Following Ref. \cite{Mas1}, consider the following SLOCC operation:
\begin{equation}  
\rho \otimes \sigma \mapsto A \otimes B (\rho \otimes \sigma) A^{\cal y} \otimes B^{\cal y},
\end{equation}
where
\begin{equation}
A = \bra{\phi_{A_{1}A_{2}}} \otimes \id_{A_{3}}, \hspace{0.3 cm} B = \bra{\phi_{B_{1}B_{2}}} \otimes \id_{B_{3}}.
\end{equation}
Here $\ket{\phi_{A_{1}A_{2}}}$ is maximally entangled state acting on the Hilbert space ${\cal H}_{A_1} \otimes {\cal H}_{A_2}$ and $\id_{A_3}$ is the identity matrix acting on the Hilbert space ${\cal H}_3$. It can be easily checked that
\begin{equation} \label{tel}
tr[A \otimes B (\rho \otimes \sigma)A^{\cal y} \otimes B^{\cal y}Z] \propto tr[\rho (\sigma^{T} \otimes Z)],
\end{equation}
for every matrix $Z \in {\cal B}({\cal H}_{A_3}\otimes {\cal H}_{B_3})$. 

We then form the LOCC operation ${\Lambda}$ as follows: we try to implement the SLOCC map defined above. This happens with probability $tr( A \otimes B (\rho \otimes \sigma) A^{\cal y} \otimes B^{\cal y})$. If we fail in applying it, something which happens with probability $1 - tr( A \otimes B (\rho \otimes \sigma) A^{\cal y} \otimes B^{\cal y})$, we through away the state obtained and prepare by LOCC a separable state $\ket{\psi}$ such that $|\braket{\psi}{\phi_{d}}|^{2} = 1/d$. From Eq. (\ref{WWW}) we see that Eq. (\ref{desiq}) is equivalent to the condition $tr(\Lambda(\rho \otimes \sigma)\phi_{d}) > 1/d$, which in turn can be written as
\begin{equation}
1/d - tr[\rho \otimes \sigma \Lambda^{\cal y}(\phi)] = tr[\rho \otimes \sigma \Lambda^{\cal y}(\id/d - \phi)] < 0,
\end{equation}
where we have used that ${\Lambda}^{\cal y}$ is unital. In turn, from Equation (\ref{tel}) one finds that the later is equivalent to
\begin{equation}
tr(\rho(\sigma^{T}\otimes (\id/d - \phi_{d}))) < 0.
\end{equation}
The Lemma then follows from Ref. \cite{Mas1}, where it was shown that for every entangled state $\sigma$, there is a $\rho \in {\cal M}({\cal H})$ satisfying the above inequality. 
\end{proof}

It is clear that it suffices to show that for every entangled state $\sigma$, 
\begin{equation} \label{Rmen}
R(\sigma) \leq - \inf_{\rho \in {\cal M}({\cal H})} \tr(W_{\rho, \Lambda} \sigma),
\end{equation}
where $\Lambda$ from now on will denote the LOCC map defined in the proof of Lemma 1. Define the set
\begin{equation}
{\cal A} := \{ W : W = \id - d \tr_{A_1B_1}[\id \otimes \rho \Lambda^{\cal y}(\phi_d)], \hspace{0.2 cm} \rho \in {\cal M}({\cal H}) \},
\end{equation}
and its associated cone:
\begin{equation}
\text{cone}({\cal A}) = \{ W : W = \sum_{i}\lambda_{i}W_{i}, \hspace{0.2 cm} \lambda_{i} 
\geq 0, \hspace{0.1 cm} W_{i} \in {\cal A} \}.
\end{equation}
From Lemma 1, we find that for every entangled state $\sigma$, there is a $W \in {\cal A}$ such that $\tr(W \sigma) < 0$.

Consider the following optimization problem:
\begin{equation} \label{optB}
\inf_{W \in {\cal B}} \tr(W \sigma),
\end{equation}
where ${\cal B} := \{ W \in \text{cone}({\cal A}) : \tr(W \pi) \leq 1, \hspace{0.1 cm} \forall \hspace{0.1 cm} \pi \in {\cal S} \}$. From the convexity of the set ${\cal M}({\cal H})$, it follows that
\begin{equation}
\text{cone}({\cal A}) = \{ W : W = \lambda W_{\rho, \Lambda}, \hspace{0.2 cm} \lambda \geq 0, \hspace{0.1 cm} \rho \in {\cal M}({\cal H}) \}.
\end{equation}
Moreover, from the definition of $G(\rho, \Lambda)$, we have for every $\rho$,
\begin{equation}
\max_{\pi \in {\cal S}} \tr(W_{\rho, \Lambda} \pi) = 1.
\end{equation}
Hence, we find that the optimization problem given by Eq. (\ref{optB}) is equivalent to the one defined on the R.H.S. of Eq. (\ref{Rmen}). 

In order to complete the proof, we show the following Lemma:

\begin{lemma}
\begin{equation} \label{optF}
R(\sigma) \leq - \inf_{W \in {\cal B}} \tr(W \sigma).
\end{equation}
\end{lemma}
\begin{proof}

The quantity in R.H.S. of Eq. (\ref{optF}) is a convex optimization problem in the variable $W$, as it consists of the minimization of a linear functional subject to the following two convex constraints:
\begin{equation*}
\tr(W \pi) \leq 1, \hspace{0.1 cm} \forall \hspace{0.1 cm} \pi \in {\cal S}, \hspace{0.2 cm} \text{and} \hspace{0.3 cm} tr(WX) \geq 0 \hspace{0.2 cm} \forall \hspace{0.1 cm} X \in {\cal A}^*.
\end{equation*}
Here, ${\cal A}^*$ is the dual cone of ${\cal A}$, defined as ${\cal A}^* = \{ X : \tr(XY) \geq 0 \hspace{0.2 cm} \forall \hspace{0.05 cm} Y \in {\cal A} \}$. Following Ref. \cite{Boyd}, let us form the Lagrangian of the problem: 
\begin{eqnarray*}
L(W, g(X), f(\pi)) = Tr(W\sigma) &-& \int\limits_{\pi \in {\cal S}} f(\pi)Tr((\id - W) \pi)d\pi \\ &-& \int\limits_{X \in {\cal A}^*} g(X)Tr(WX)dX, \nonumber
\end{eqnarray*}
where $f(\pi)$ and $g(X)$ are the Lagrange multipliers (in this case given by functions or, more precisely, generalized functions). The dual problem is then
\begin{eqnarray} \label{dual22}
\quad \mbox{maximize} \quad & - \tr(\displaystyle\int\limits_{\pi \in {\cal S}} f(\pi)\pi d\pi)  \\ 
\quad \mbox{subject to} \quad & f(\pi) \geq 0, \hspace{0.2 cm} \forall \pi \in {\cal S}  \nonumber \\ 
 & g(X) \geq 0, \hspace{0.2 cm} \forall X \in {\cal A}^{*} \nonumber \\
 & \sigma + {\displaystyle \int\limits_{\pi \in {\cal S}}} f(\pi)\pi d\pi = {\displaystyle \int\limits_{X \in {\cal A}^{*}}} g(X)X dX. \nonumber
\end{eqnarray} 
As $f(\pi) \geq 0$ and $g(X) \geq 0$, we have that
\begin{equation}
\pi' :=  {\displaystyle \int\limits_{\pi \in {\cal S}}} f(\pi)\pi d\pi
\end{equation}
is a non-normalized separable state and that
\begin{equation}
X' := {\displaystyle \int\limits_{X \in {\cal A}^{*}}} g(X)X dX \hspace{0.2 cm} \in {\cal A}^*.  
\end{equation}
It then follows that the optimal solution of Eq. (\ref{dual22}) is $-s$, where $s$ is the minimal number such that
\begin{equation} 
\sigma + s\pi = (1 + s)X,
\end{equation}
where $\pi := \pi' / \tr(\pi')$ and $X := X'/\tr(X')$. As $\sigma$ and $\pi$ are positive semi-definite, we find that $X$ is a state. Since, by Lemma 1, every entangled state is detected by an entanglement witness belonging to ${\cal A}$, it follows that $X$ is  a separable state. This in turn implies that $s$ is larger or equal to $R(\sigma)$. It is easy to check that Slater's conditions \cite{Boyd} apply to optimization problem (\ref{dual22}), so that the primal problem (given by the R.H.S. of Eq. (\ref{optF})) and its dual (described by Eq. (\ref{dual22})) have the same optimal solutions. This finishes the proof of this Lemma and of Theorem 1. 
\end{proof}

In this work we considered the question of how to quantify the activation capacity of bi-partite entangled states. It was show that this capacity, when quantified by a particular figure of merit, turns out to be the robustness of entanglement of the state in question. This connection between activation properties and the robustness of the quantum correlations contained in the state, interesting on its own, gives new insights into a number of question in the theory of entanglement. In particular, we find the robustness as the first geometric entanglement measure with an operational meaning and the only known operational entanglement measure which is non-zero on every entangled state. This, taken together with fact that the robustness can be systematically approximately calculated both from the top and bottom by efficient procedures \cite{Bran1, Doh1, Eis1}, indicates that the robustness might be a very suitable measure in the study of entanglement, and in particular in the elucidation of the properties of bound entanglement \cite{Hor2}. Finally, the expectation values of entanglement witnesses are given a new interpretation as lower bounds to the performance of the measured state in a particular quantum information task - more precisely, a lower bound to its activation capacity. 
	
The author is thankful to K. Audenaert, D. Cavalcanti, J. Eisert, P. Hyllus, Ll. Masanes, M. Plenio, M. Terra Cunha, R. Vianna, and S. Virmani for discussions and suggestions on the manuscript. This work has been supported by the Brazilian agency Conselho Nacional de Desenvolvimento Cient\'ifico e Tecnol\'ogico (CNPq).

\end{document}